\documentclass{PoS}
\usepackage{nico}
\usepackage{subfigure, epsfig, rotate, epsf,  amssymb, amsxtra, placeins, url }

\ShortTitle{NP running and renormalization of kaon 4-quark operators
}

\title{Non-perturbative running and renormalization 
of kaon four-quark operators
with $n_f=2+1$ domain-wall fermions}

\newcommand\edinb{Tait Institute, University of Edinburgh, Edinburgh EH9 3JZ, UK}
\newcommand\soton{School of Physics and Astronomy, University of
  Southampton,  Southampton SO17 1BJ, UK}

\author{P.~A.~Boyle, N.~Garron$^*$   \\ 
  \edinb}

\author{A.~T.~Lytle$^*$\\
  \soton}

\author{ RBC-UKQCD collaborations  \\
  \hfill \it Edinburgh 2011/31 \\ 
  \hfill \it SHEP-1135 }

\ph{\speaker{N.~Garron and A.~T.~Lytle}}

\abstract{
We compute the renormalization factors of four-quark operators 
needed for the study of $K\to\pi\pi$ decay in the $\Delta I=3/2$ channel. 
We evaluate the Z-factors at a low energy scale ($\mu_0=1.145 \GeV$) 
using four different non-exceptional RI-SMOM schemes 
on a large, coarse lattice ($a\sim 0.14\fm$) on which the bare matrix elements are also computed.
Then we compute the universal, non-perturbative, scale evolution matrix 
of these renormalization factors between $\mu_0$ and $3\GeV$.
We give the numerical results for the different steps of the computation
in two different non-exceptional lattice schemes, and 
the connection to $\msbar$ at $3\GeV$ is made using one-loop perturbation theory.
}

\FullConference{ The XXIX International Symposium on Lattice Field Theory - Lattice 2011\\
July 10-16, 2011\\
Squaw Valley, Lake Tahoe, California}

\begin{document}
\section{Introduction}
\label{sec:intro}
The RBC-UKQCD collaborations have recently achieved the computation 
of the $\Delta I=3/2$ part of $K\to\pi\pi$ decays~\cite{Blum:2011ng,bob:lat11}.
At leading order of the operator product expansion, there are three operators 
that enter the computation : a tree level operator $Q_1^{3/2}$ and 
two electroweak penguins $Q_7^{3/2}$ and $Q_8^{3/2}$ 
\bea
Q_1^{3/2} &=&
 (\bar s_i \gmuL d_i) 
\big[ 
(\bar u_j  \gmuL u_j) - (\bar d_j  \gmuL d_j)) 
\big]
+( \bar s_i \gmuL u_i)(\bar u_j \gmuL d_j)) 
\,,
\\
Q_7^{3/2} &=& 
(\bar s_i \gmuL d_i) 
\big[ 
(\bar u_j  \gmuR u_j) - (\bar s_j  \gmuR s_j)) 
\big]
+( \bar s_i \gmuL u_i)(\bar u_j \gmuR d_j))
\,,
\\
Q_8^{3/2} &=& 
(\bar s_i \gmuL d_j) 
\big[ 
(\bar u_j  \gmuR u_i) - (\bar s_j  \gmuR s_i)) 
  \big]
+( \bar s_i \gmuL u_j)(\bar u_j \gmuR d_i))
\,,
\eea
where $\gamma_\mu^{R,L}=\gamma_\mu(1\pm\gamma_5)$ and $i,j$ are colour indices.
In order to 
simulate a 2-hadron final state with (nearly) physical kinematics and quark
masses, 
the computation of the matrix elements $\la\pi\pi|Q^{3/2}_i|K\ra$ 
was done on a large volume (of space extent $L_{0}\sim 4.6 \text{ fm}$),
with a rather coarse lattice spacing $a_0\sim0.14 \fm$
(we refer to this lattice as IDSDR: 
Iwasaki with Dislocation Suppressing Determinant Ratio, see~\cite{kelly:lat11} for more
details). 
The extraction of the bare matrix elements has been reported 
first last year~\cite{Goode:2011kb} and updated this year~\cite{Goode:2011pd}.
In this work we explain our method to renormalize non-perturbatively 
the bare matrix elements computed on this lattice. 
Since the lattice spacing is rather coarse, the usual Rome-Southampton 
condition~\cite{Martinelli:1994ty} does not hold:
in the region where the discretisation effects are under control
$\Lambda_{\rm QCD}^2 \sim \mu_0^2\ll (\pi /a_0)^2$. 
This problem is circumvented by the use of a step-scaling 
matrix~\cite{Arthur:2010ht,Arthur:2011cn}. 
Our strategy involves three different steps: 
\begin{itemize}
\item[1.]
We evaluate the Z-factors at low energy $\mu_0$ on the IDSDR 
lattice using four different RI-SMOM schemes and compute 
the relevant renormalized matrix elements.
The scale $\mu_{\rm 0}$ is 
such that the associated discretisation errors are small
but, compared to the Rome-Southampton window, we do not require the non-perturbative effects to be small.
Instead one just has to ensure that the finite volume effects are negligible , 
so the renormalization window becomes 
$$L_0^{-2} \ll \mu_0^2 \ll (\pi/a_0)^2\;.$$ 
\item[2.]
We compute the scale evolution between $\mu_0$ and $\mu=3 \GeV$
of these operators on finer lattices\footnote{We refer to these as
the IW(asaki) lattices.}
(in practice we use $a\sim 0.086 \fm, 0.114 \fm $), 
on which the high scale lies in the usual Rome-Southampton window 
$$
\Lambda_{\rm QCD}^{2} \ll \mu^2 \ll (\pi/a)^2 \;.
$$
We extrapolate the result to the continuum and obtain the universal running in this energy range 
for the different renormalization schemes.
\item[3.]
At the scale $\mu=3 \GeV$ we convert the results to $\msbar$ using 
one-loop perturbation theory~\cite{Lehner:2011fz}.
\end{itemize}
\begin{figure}
\centering
\includegraphics[width=\textwidth]{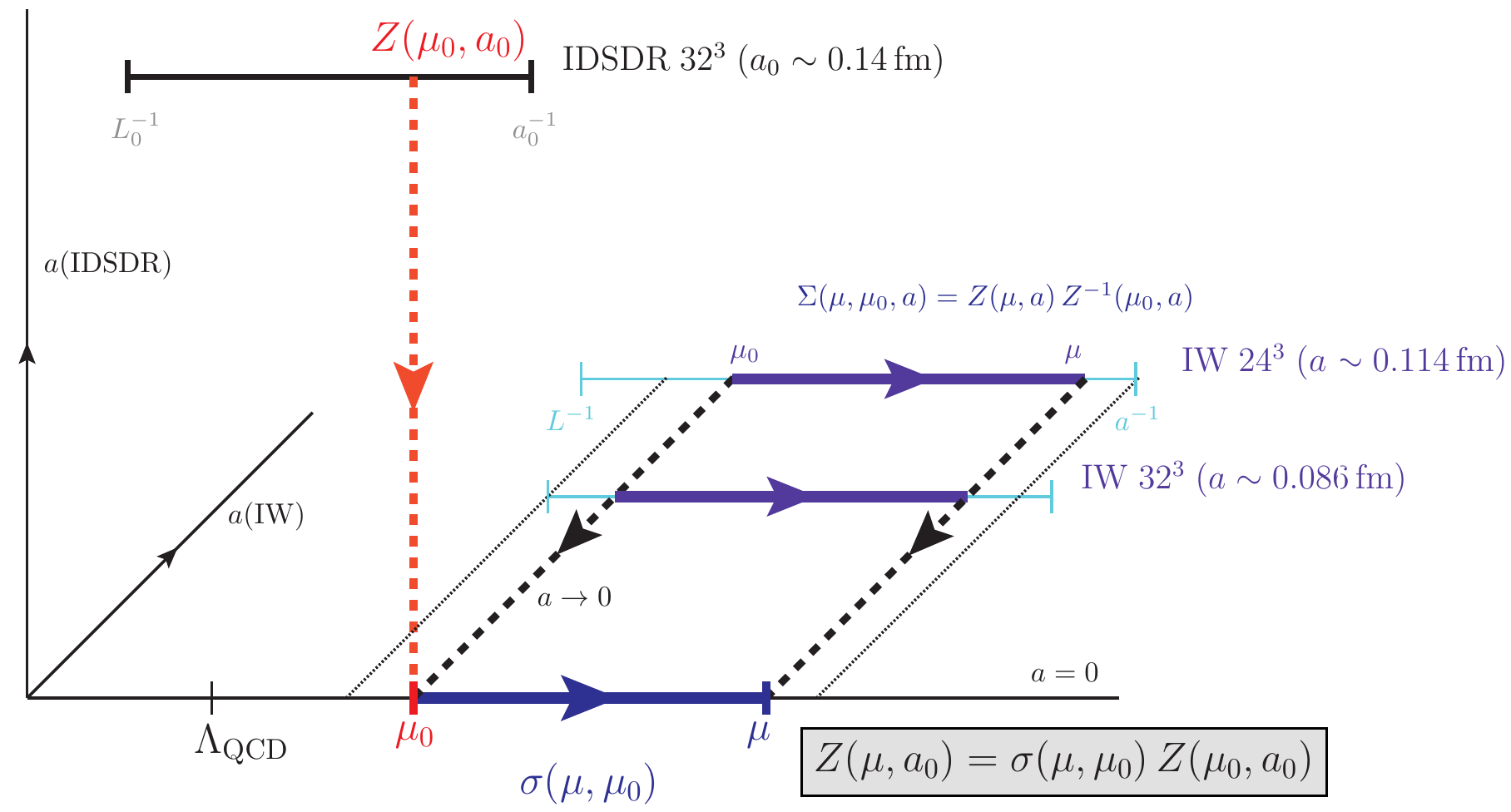}
\caption{Strategy of the renormalization procedure. 
The horizontal axis represents the energy scale and the two other axes
represent the lattice spacings of the IW and IDSDR ensembles.
As explained in the text, we compute the renormalization matrix 
$Z(\mu_0,a_0)$ at low energy $\mu_0$ on the coarse IDSDR lattice. 
There, the usual Rome-Southampton condition $\Lambda_{\rm QCD} \ll \mu_0 \ll a_0^{-1}$ 
does not hold. Thus we combine this result with the continuum non-perturbative 
scale evolution $\sigma(\mu,\mu_0)$ 
extracted from two finer IW lattices and obtain the renormalization 
factors at a perturbative scale $\mu$, where we match to $\msbar$.
On each set of lattices the straight line represents (symbolically) the accessible 
energy range $L^{-1} \ll \mu \ll a^{-1}$, where both finite volume and discretisation
effects are under control.}
\label{fig:strategy}
\end{figure}
Our strategy - depicted in Fig.~\ref{fig:strategy} -
can be summarised by the following equation:
\be
\label{eq:main}
\langle O^{\overline{\rm MS}}(\mu) \rangle
=
C^{\msbar\leftarrow \schemeS}(\mu) \times
\underbrace{\sigma^{\schemeS}(\mu,\mu_0)}_{\mbox{Fine lattices} \;  a\to 0}
\times \underbrace{Z^{\schemeS}(\mu_0)\langle O^{bare}(\mu_0) \rangle}_
{\mbox{Coarse lattice }}
\;,
\ee
where $C^{\overline{\rm MS}\leftarrow \schemeS}(\mu)$ represents the matrix of matching factors which converts
the $Z$-matrix computed in the scheme $\schemeS$ to the scheme $\overline{\rm MS}$, and 
$\sigma^{\schemeS}(\mu,\mu_0)$ is the non-perturbative running matrix in the scheme $\schemeS$
(see Section~\ref{sec:stepscaling} for a precise definition).
Since at the moment we have only one lattice spacing on the IDSDR lattice,
the previous equation will be affected by lattice artefacts (which are estimated), 
but in principle we could take the continuum limit of $Z^{\schemeS}(\mu_0)\langle O^{bare}(\mu_0) \rangle$ 
and interpret Eq.~\eqref{eq:main} as a continuum equation.
The final result of Eq.~(\ref{eq:main}) does not depend on the 
choice of intermediate schemes $\schemeS$ (up to 
truncation errors in perturbation theory). We use four different 
non-exceptional schemes (but as motivated in~\cite{Aoki:2010pe} we focus only on two schemes)
and use the difference to estimate 
the size of these errors.\\

The remainder of the text is organised as follows: in the next section we 
briefly review the RI-MOM renormalization procedure used in our calculations,
focusing on recent innovations extending the original 
proposal of~\cite{Martinelli:1994ty}.
Section~\ref{sec:schemes} defines precisely our renormalization prescriptions.
Section~\ref{sec:Zmu0} presents our results for the renormalization
factors at low energy on the IDSDR lattice in two different schemes.
In Section~\ref{sec:stepscaling} we define the step-scaling function matrix 
which gives the scale evolution of the operators under considerations and 
give our numerical results for these two schemes.
Final results in the $\msbar$ scheme are presented in Section~\ref{sec:msbar}, 
and our conclusions in Section~\ref{sec:conclusion}.

\section{Background on RI-MOM}
\label{sec:baground}
We impose renormalization conditions
directly on the lattice using RI-MOM type schemes, as first proposed in~\cite{Martinelli:1994ty}.
Because the correlation functions are computed using quarks with fixed external momenta,
it is advantageous to compute quark propagators using 
momentum sources~\cite{Gockeler:1998ye}, $\eta \sim e^{ipx}$.
This allows the free spatial index of the propagator to be summed at the vertex,
greatly improving the signal and resulting in very small statistical errors, 
even when using relatively few configurations.\\

We use a modified kinematic setup to that originally proposed in~\cite{Martinelli:1994ty},
called ``non-exceptional"  kinematics, 
in contrast to the original ``exceptional" configuration.
Instead of using a single external momentum $p$ with zero momentum inserted at the vertex,
one uses quark propagators with external momenta $p_{1}$ and $p_{2}$ satisfying
$p_{1}^{2} = p_{2}^{2} = (p_{1}-p_{2})^{2}$,
with $(p_{1} - p_{2})$ inserted at the vertex.
This maintains a single renormalization scale, but suppresses channels in which no momentum
is flowing, resulting in a stronger suppression of chiral symmetry breaking effects.
Details may be found in~\cite{Aoki:2007xm}.\\

Finally, we make use of twisted boundary conditions~\cite{Arthur:2010ht}, 
in which the quark fields pick up an arbitrary phase at the boundary of the lattice.
In this way, we can compute quark propagators at arbitrary momentum instead of only
at discrete Fourier modes, i.e. if $q$ represents the quark field then
\begin{equation}
q(x+L) = e^{i \theta} q(x), \quad p = \frac{2 \pi}{L} n + \frac{\theta}{L} \, .
\end{equation}
There are several reasons for using twisted boundary conditions. 
The principal motivation is that the direction of $p$ can be chosen to remain constant as its magnitude 
and the lattice spacing are changed.
{\it This implies the existence of the continuum limit of the non-perturbative running 
for all the different momenta.}
For a given discretization, all lattice artifacts have a fixed parametric dependence 
on the renormalization scale.
Not only will the resultant data be very smooth in $(ap)^{2}$, it means that data from different 
lattice spacings lie along a continuum trajectory.
They are also practical advantages: we can simulate the same 
physical scale $\mu_0$ on both the IDSDR and the IW ensembles (as long as we know 
the lattice spacings with sufficient precision). Also we do not need 
to subtract perturbatively the O(4)-lattice artefacts, or 
to give any more-or-less arbitrary prescription to choose our momenta.
Thus the use of twisted boundary conditions forms an essential part of our calculation.

\section{Renormalization in RI-SMOM schemes}
\label{sec:schemes}
In  \cite{Aoki:2010pe}, $B_K$ was renormalized in four different RI-SMOM schemes.
Here we generalise this procedure to the case of operator mixing. 
The operator $Q_1$ belongs to the $(27,1)$ representation of 
$SU(3)_{L} \times SU(3)_{R}$, whereas $Q_7$ and $Q_8$ belong to $(8,8)$.
Thus, if chiral symmetry is realised, the renormalization pattern is the following
(in order to simplify the notations we drop the superscript~$3/2$):
\bea
Q_1^R  &=&  Z_{(27,1)} \; Q_1^{bare} 
\\
& \nn
\\
\left(
\begin{array}{c}
Q_7^R\\
Q_8^R
\end{array}
\right)
&=&
Z_{(8,8)}
\left(
\begin{array}{c}
Q_7^{bare}\\
Q_8^{bare} 
\end{array}
\right)
=
\left(
\begin{array}{cc}
Z_{77}  &  Z_{78}    \\
Z_{87}  &  Z_{88}  
\end{array}
\right)
\;
\left(
\begin{array}{c}
Q_7^{bare}\\
Q_8^{bare} 
\end{array}
\right)
\;.
\eea
Moreover the renormalization factors of these operators are related to those
of $\Delta S=2$ operators relevant for neutral kaon mixing within and beyond the Standard Model,
which have been already studied on the lattice 
(see e.g. \cite{Babich:2006bh,Boyle:2011kn,Dimopoulos:2010wq,Wennekers:2008sg}).
For example $Z_{(27,1)}$ is the same as $Z_{VV+AA}=Z_{B_K}\,Z_A^2$ of \cite{Aoki:2010pe}.
For completeness we repeat here some details of this 
computation: first we consider 
the process
\be
d(p_1)\bar{s}(-p_2)\to\bar{d}(-p_1)u(p_2)
\ee
with $p_1^2=p_2^2=(p_1-p_2)^2=\mu^2$ for a variety of momenta satisfying this condition. 
We call 
$\Lambda_{\alpha \beta, \gamma \delta}^{ij, kl}$ the corresponding amputated Green 
function evaluated on Landau gauge-fixed configurations (the colour indices $i,j,\ldots$
and Dirac indices $\alpha,\beta,\ldots$ correspond to the external states). 
We then have to project this Green function onto its Dirac-colour structure, but since 
$\mu\ne0$ the choice of the projector is not unique. We define two projectors
($N_C$ is the number of colours):
\bea
P^{(\gamma^{\mu})\,ij, kl}_{\alpha \beta, \gamma \delta} 
&=&\frac{1}{128N_c(N_c+1)}\,
\left[ (\gmuL)_{\beta\alpha} (\gmuL)_{\delta\gamma} \right]
 \delta^{ij} \delta^{kl}
\\
P^{(\qslashs)\,ij, kl}_{\alpha \beta, \gamma \delta} 
&=& \frac{1}{32q^{2}N_c(N_c+1)}
\left[ (\qslashL)_{\beta\alpha} (\qslashL)_{\delta\gamma} \right]
\delta^{ij} \delta^{kl} \,.
\eea
which act on $\Lambda$ in the following way:
\be
\label{eq:proj}
M\equiv P\{\Lambda\} \equiv P^{ij, kl}_{\alpha \beta, \gamma \delta} \Lambda_{\alpha \beta, \gamma \delta}^{ij, kl}
\ee
To renormalize the quark field we use two schemes: the $\gamma_\mu$-scheme
and the $\qslash$-scheme:
\bea
Z_q^{(\qslashs)}=\frac{q^\mu}{12q^2}\,\mathrm{Tr}[\Lambda_V^\mu\qslash] \\
Z_q^{(\gamma_\mu)}=\frac{1}{48}\,\mathrm{Tr}[\Lambda_V^\mu\gamma^\mu]\,,
\eea
where $\Lambda_V^\mu$ is the amputated Green function of the conserved vector current.
The renormalization factor $Z_{(27,1)}^{(A,B)}$ in the scheme ${\cal S}=(A,B)$ is then obtained 
by imposing
\be
Z_{(27,1)}^{(A,B)} = 
(Z^{(B)}_q)^2 \big[ P^{(A)}\{\Lambda\} \big]^{-1}
\ee
where $A$ and $B$ can be either $\gamma^{\mu}$ or $\qslash$, 
in this way we have defined four different non-exceptional RI-SMOM
schemes.\\

For the electroweak penguins we generalise the previous equations 
to the operator mixing case: they are now two different vertex functions
$\Lambda_7$ and $\Lambda_8$ and two projectors $P_7$ and $P_8$
given by 
\bea
\left[P^{(\gamma^{\mu})}_{7}\right]^{ij, kl}_{\alpha \beta, \gamma \delta} &=& 
\left[ (\gmuL)_{\beta\alpha} (\gmuR)_{\delta\gamma}  \right]
 \delta^{ij} \delta^{kl}\\
\left[P^{(\gamma^{\mu})}_{8}\right]^{ij, kl}_{\alpha \beta, \gamma \delta} &= &
\left[ (\gmuL)_{\beta\alpha} (\gmuR)_{\delta\gamma}  \right]
\delta^{il} \delta^{jk}
\eea 
in the $\gamma^{\mu}$ scheme and 
\bea
\left[P^{(\qslashs)}_{7}\right]^{ij, kl}_{\alpha \beta, \gamma \delta} 
&=& \frac{1}{q^{2}}\left[ (\qslashL)_{\beta\alpha} (\qslashR)_{\delta\gamma} \right]\delta^{ij} \delta^{kl} \\
\left[P^{(\qslashs)}_{8}\right]^{ij, kl}_{\alpha \beta, \gamma \delta} 
&=& \frac{1}{q^{2}}\left[ (\qslashL)_{\beta\alpha} (\qslashR)_{\delta\gamma} \right]  \delta^{il} \delta^{jk} \, .
\eea
in the $\qslash$ scheme. Let us call $M$ the matrix defined by 
\be
\label{eq:proj_mix}
M^{(A)}_{ij} = P^{(A)}_j\{\Lambda_i\}\;, \qquad (i,j=7,8) \;,
\ee
where the projector acts in the same way as in Eq.~\eqref{eq:proj}.
The two by two renormalization matrix $Z^{(A,B)}$ in the scheme
${\cal S} = (A,B)$ is then defined by
\be
Z^{(A,B)} = 
(Z^{(B)}_q)^2 \, F \, \left[M^{(A)}\right]^{-1} \;,
\ee
where $F$ is the tree-level value of the matrix $M$.

\section{Computation of the Z-factors at low energy}
\label{sec:Zmu0}
As explained in the Introduction, since the extraction of the bare matrix elements 
is done on the coarse ISDSR lattice, the first step consists in 
computing the renormalization factors at low energy ($\mu_0 = 1.145 \GeV$) 
on the same lattice.
We follow the procedure described in Section~\ref{sec:schemes} and obtain
after a chiral extrapolation
\begin{eqnarray} 
Z^{(\gamma^{\mu}, \gamma^{\mu})}_{(27,1)} (\mu_0) &= 0.443\,(01) \,,\qquad
Z^{(\gamma^{\mu}, \gamma^{\mu})}_{(8,8)} (\mu_0) &=
\begin{pmatrix}
\ph{-}0.505\,(01) &     	-0.114\,(01)	\\
-0.022\,(03)      &    \ph{-}0.231\,(02)	\\
\end{pmatrix} \,, \\[3mm]
Z^{(\qslashs, \qslashs)}_{(27,1)} (\mu_0) &= 0.489\,(01) \,,\qquad
Z^{(\qslashs, \qslashs)}_{(8,8)}  (\mu_0) &=
\begin{pmatrix}
\ph{-}0.510\,(02)	&	-0.116\,(01) 	\\
-0.077\,(06)	&	\ph{-}0.305\,(04) 	\\
\end{pmatrix} \,,
\end{eqnarray}
where the quoted errors are statistical only.
Here and in the remainder of this paper we estimate and propagate the statistical errors 
by using 100 bootstrap samples.\\

We have done the computation for the four different non-exceptional schemes defined in section~\ref{sec:schemes},
but here and in the remainder of the text we choose to quote only the results in the 
${(\gamma^{\mu}, \gamma^{\mu})}$ and in the $(\qslash, \qslash)$-scheme. 
In Fig.~\ref{fig:Z11DSDR} we show the chiral extrapolation of the renormalization factor 
$Z_{11}=Z_{(27,1)}$
in the ${(\gamma^{\mu}, \gamma^{\mu})}$-scheme, 
evaluated at the scale $\mu_0$. 
Our setup is unitary in the light sector, but the strange quark is partially quenched: 
we consider only degenerate valence quark masses which are  set equal to the sea light 
quark masses and extrapolated to zero, whereas the sea quark mass of the strange is fixed 
(to its physical value). As a consequence our results are affected by a small 
systematic error, which was evaluated in~\cite{Aoki:2010pe} for the $(27,1)$ operator.

\section{Computation of the non-perturbative running}
\label{sec:stepscaling}
We now consider two fine Iwasaki lattices, the details of which can be found 
in~\cite{Aoki:2010dy,Allton:2008pn}.
There we compute the renormalization factors following again Section~\ref{sec:schemes}. 
In particular we obtain $M$ in the four different
RI-MOM schemes (in the following $M$ can either be a scalar -- 
see Eq.~\eqref{eq:proj} -- or a matrix -- see Eq.~\eqref{eq:proj_mix} --
depending if we consider the multiplicative 
renormalization case or the operators mixing case). 
In order to introduce the step-scaling matrix, we follow~\cite{Arthur:2011cn}:
at finite lattice spacing $a$ and for a given renormalization scale $\mu$ we consider
\be
R_{\schemeS}(\mu,a)=\lim_{m\to 0}\left[\Lambda^2_{\rm A}(\mu,a,m) \, M^{-1}(\mu,a,m)\right]
\;,
\ee
and we define the step scaling
\be
\label{eq:ssm}
\sigma^\schemeS(\mu,s\mu)=\lim_{a\to 0}\Sigma^\schemeS(\mu,s\mu,a)=\lim_{a\to 0}{\left[{R_\schemeS}(\mu,a)\times R_\schemeS^{-1}(s\mu,a)\right]}\;,
\ee
(we normalise by $\Lambda^2_{\rm A}$ in order to cancel the quark wavefunction 
renormalization).
One important point is that although the quantities $M$, $Z$ and $R$
depend on the details of the computation this is not the case for the 
step scaling matrix which has well-defined continuum limit 
and is thus universal: 
it depends only on the choice of the renormalization scheme $\schemeS$
and on the number of flavours.\\

\begin{figure}[t]
\centering
\begin{tabular}{cc}
\includegraphics[width=7.cm]{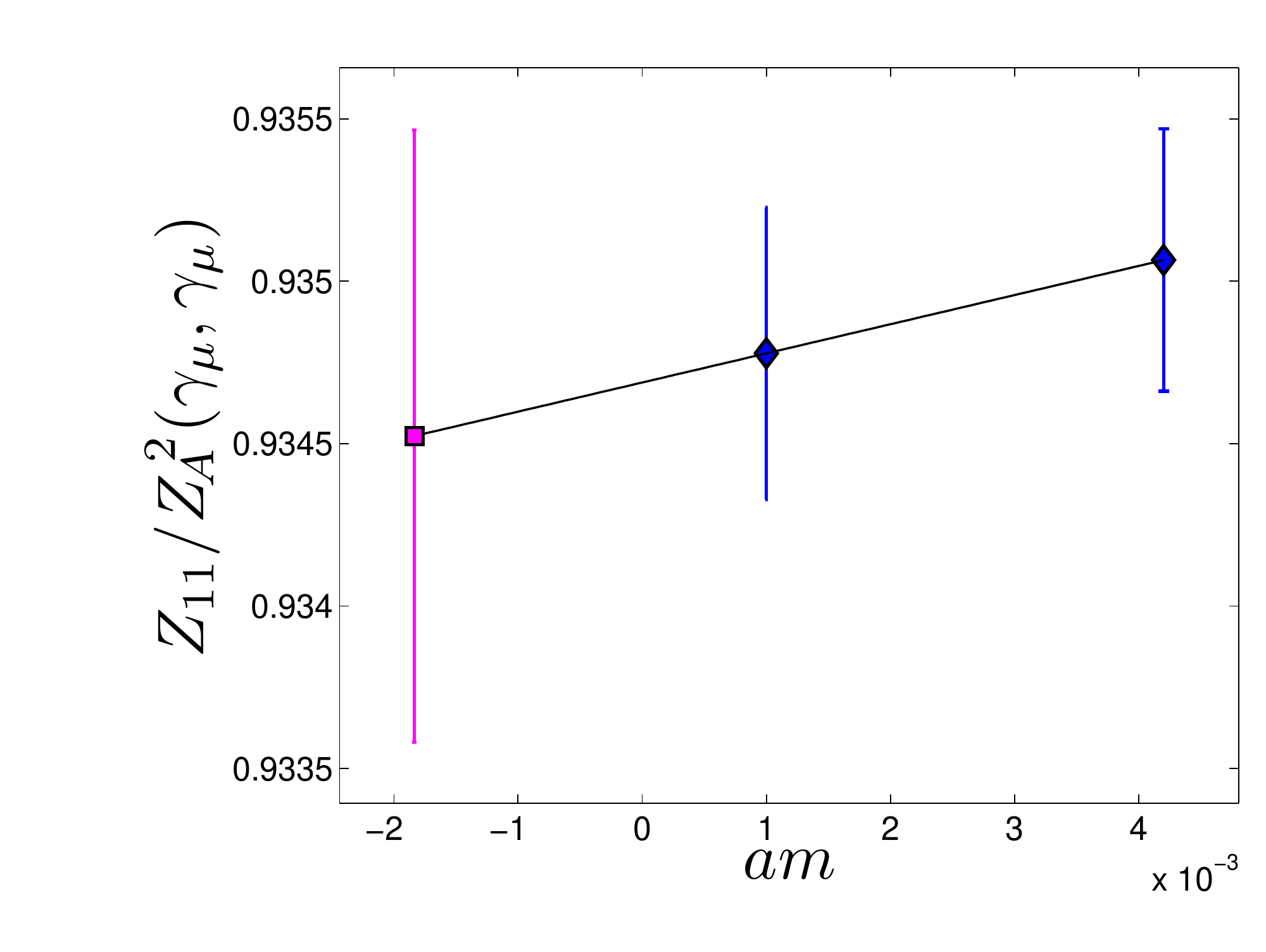} &
\includegraphics[width=7.cm]{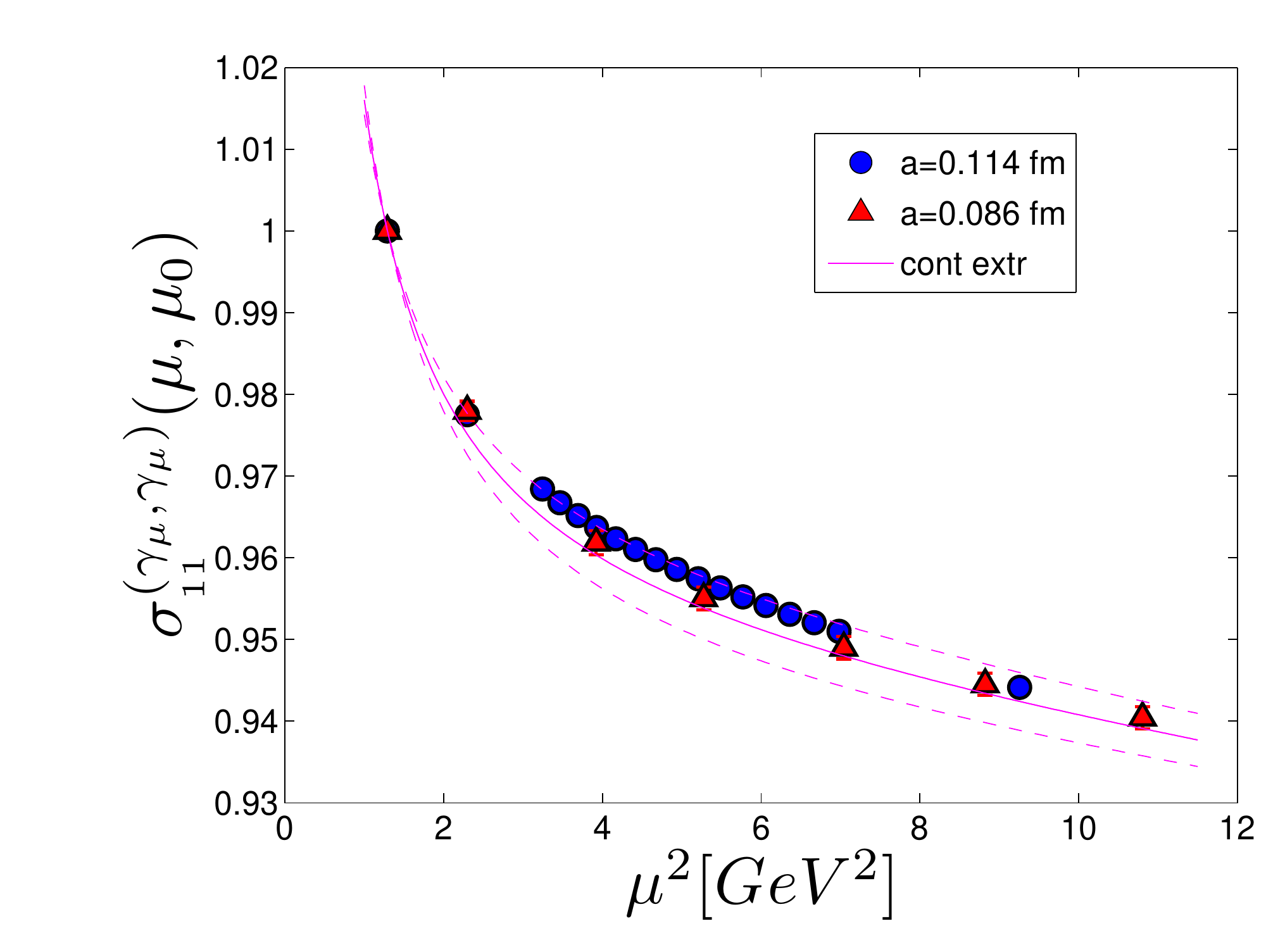}
\end{tabular}
\caption{The left plot shows the chiral extrapolation of the renormalization factor of the $(27,1)$ operator,
normalised by $Z_A^2$, 
computed on the IDSDR lattice in the $(\gamma_\mu,\gamma_\mu)$ scheme. On the x-axis,
$am$ represents the bare light quark mass.
We show the result at finite mass and in the chiral limit. 
On the right we show the non-perturbative scale evolution of the same operator
in the same scheme, at finite lattice spacing and in the continuum limit.}
\label{fig:Z11DSDR}
\end{figure}

\begin{figure}[t]
\centering
\begin{tabular}{cc}
\includegraphics[width=7.cm]{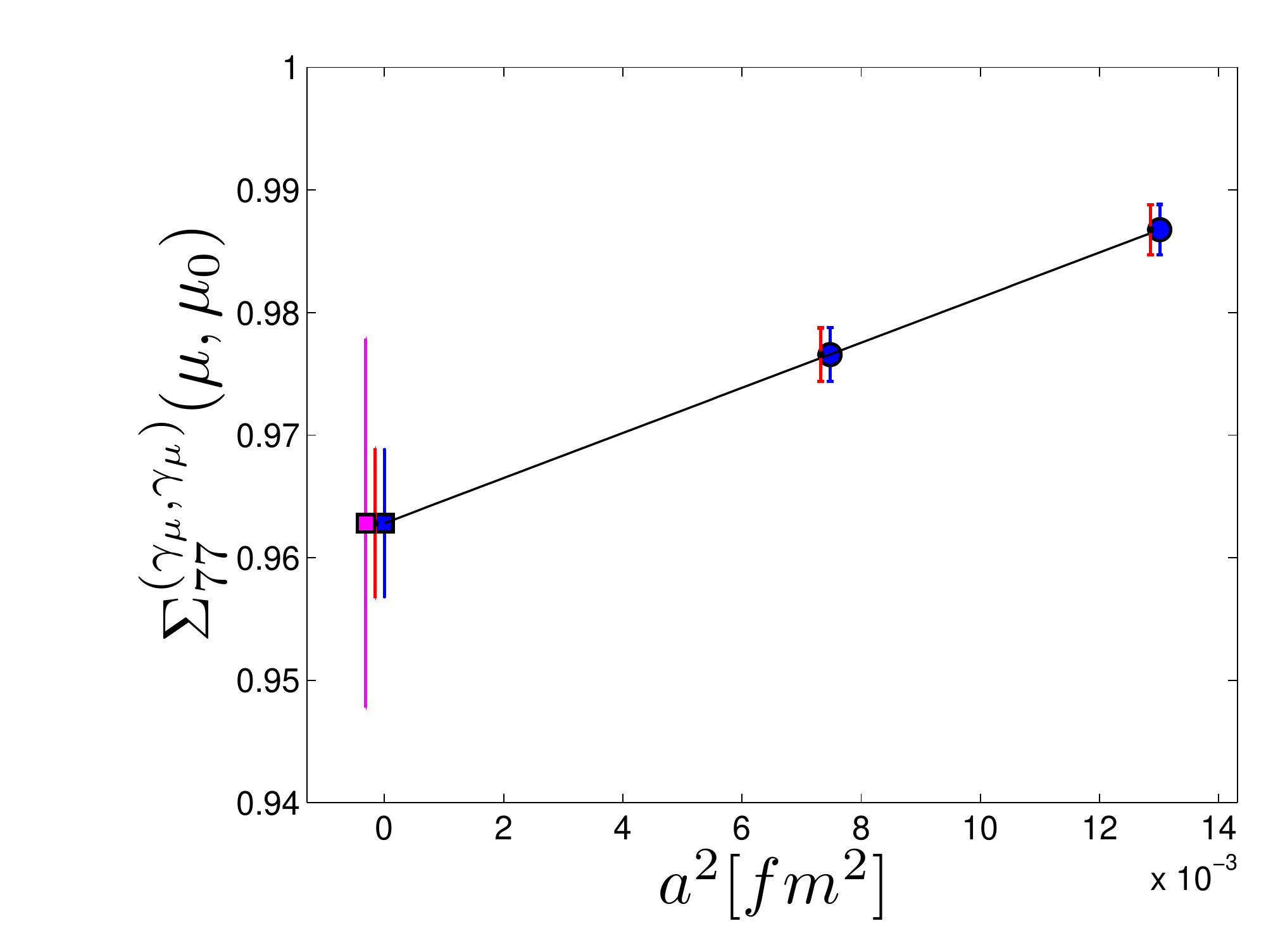} &
\includegraphics[width=7.cm]{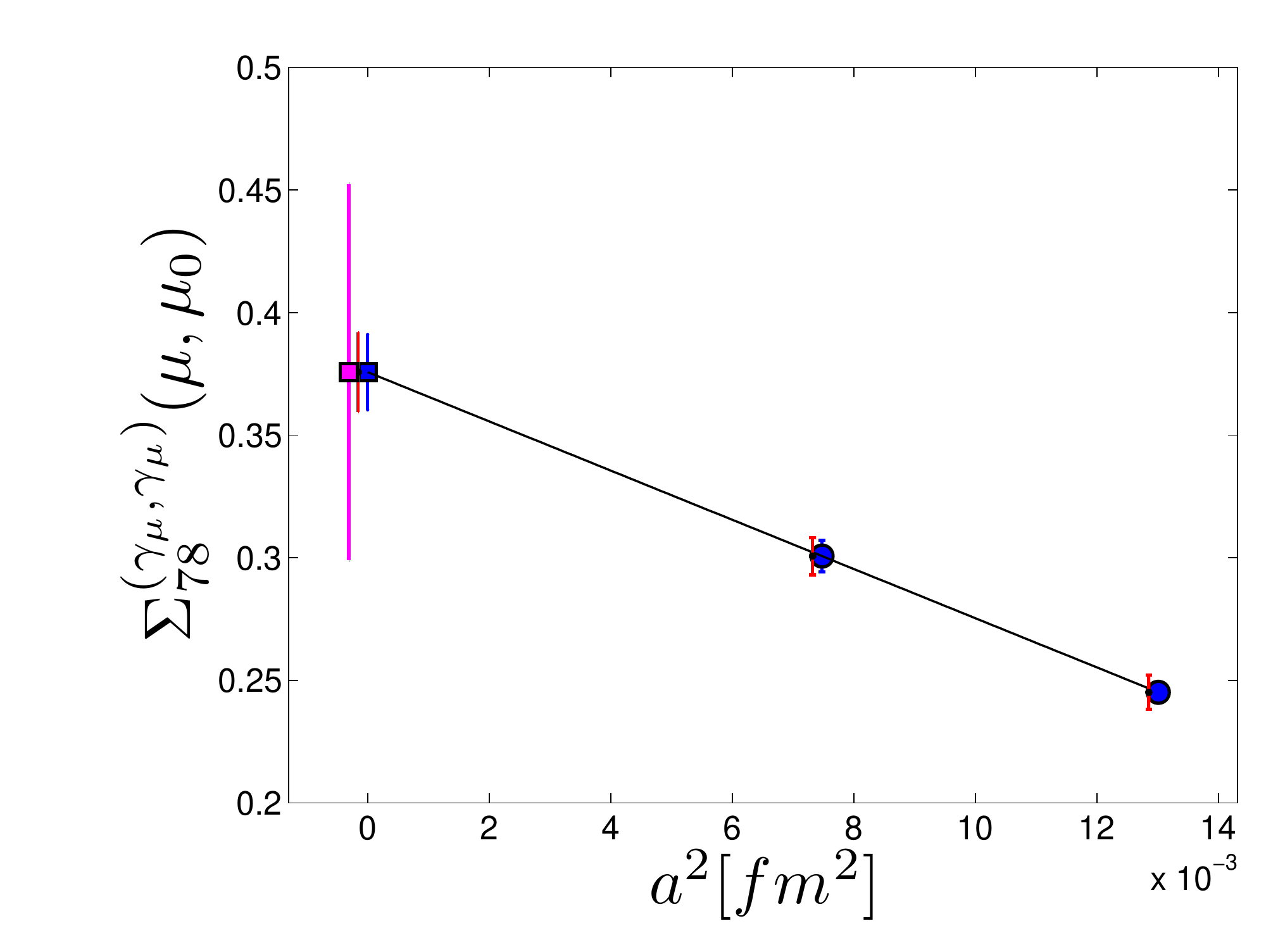}\\
\includegraphics[width=7.cm]{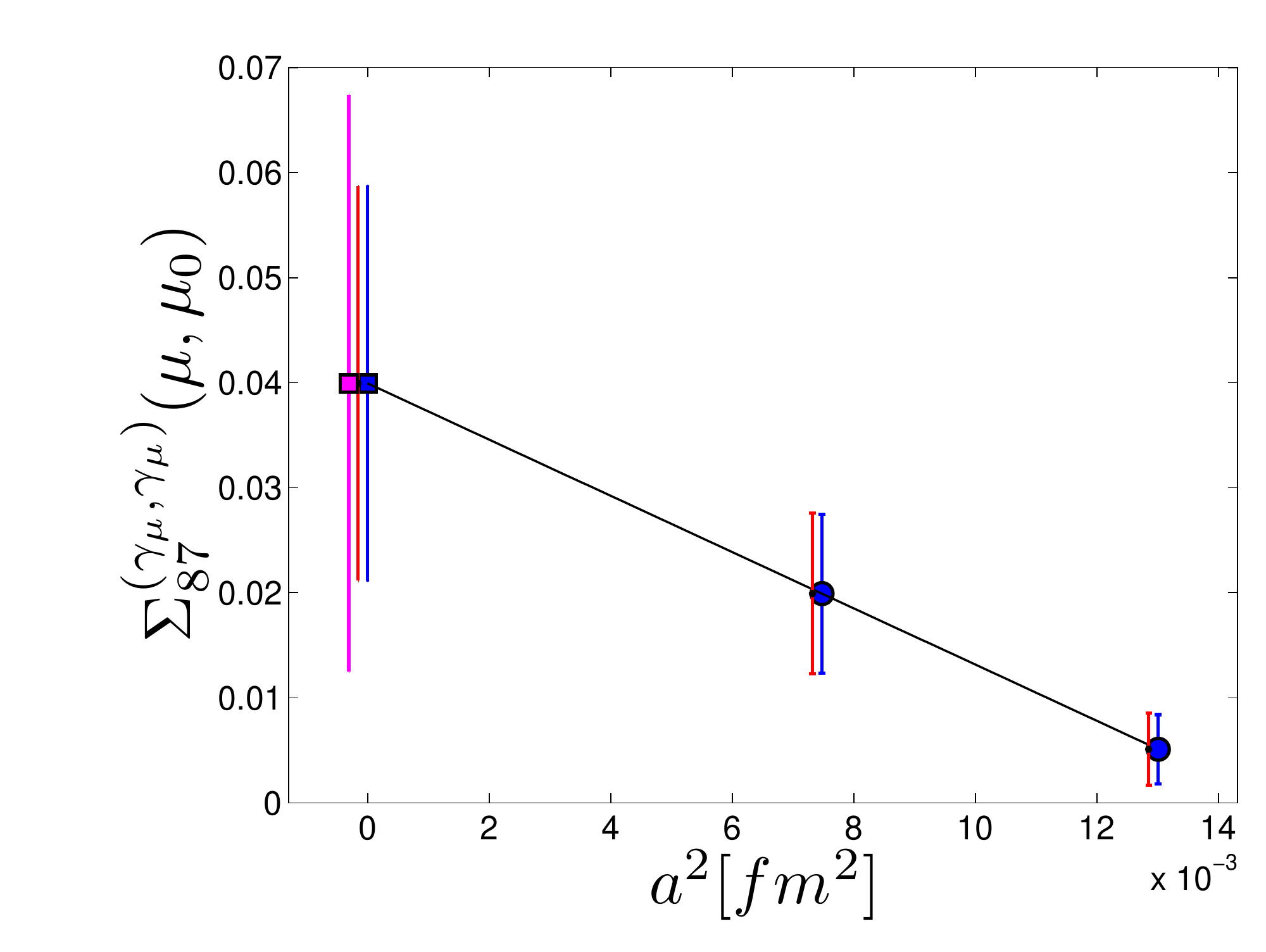}&
\includegraphics[width=7.cm]{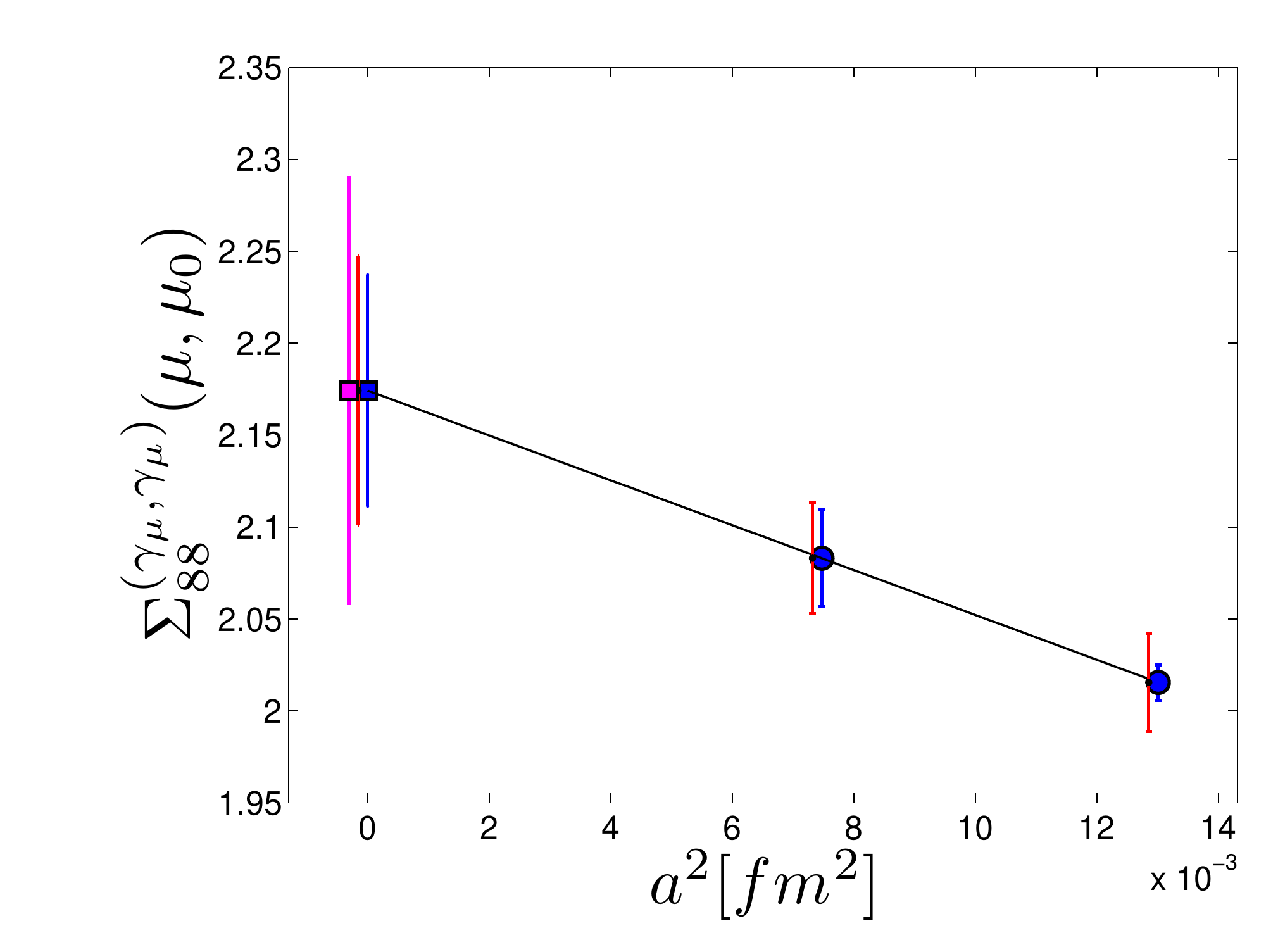}
\end{tabular}
\caption{Continuum extrapolation of the step scaling matrix elements
of the $(8,8)$ operators. At finite lattice spacing we show the central value
together with the naive statistical error. Slightly shifted to the left, we show 
the error bar where we have taken into account the error from the lattice spacing.
In the continuum we add in quadrature an error which is defined to be 
the difference between the continuum extrapolation and the finest lattice spacing;
the resulting error is displayed at the far left.}
\label{fig:sigma88}
\end{figure}

When performing the continuum extrapolation, we match scales on the different lattices
by interpolating the simulated data, which are very smooth on account of our use of twisted boundary 
conditions.  
In the right panel of Fig.~\ref{fig:Z11DSDR}, we show the step scaling function of
the $(27,1)$ operators in the $(\gamma_\mu,\gamma_\mu$)-scheme at finite lattice spacing and 
extrapolated to the continuum.
Twisted boundary conditions also ensure the data lie along a continuum trajectory, 
and with two Iwasaki ensembles we 
attempt to remove the lattice artefacts by doing a linear fit in $a^2$.
Since we have only two different lattice spacings, we choose to include a
(rather conservative) systematic error coming from the difference between 
the results on our finest Iwasaki lattice and the continuum-extrapolated results. 
As an example, we show the continuum extrapolation of the $(8,8)$ operators 
in Fig.\ref{fig:sigma88}.
With $\mu_0 = 1.145 \GeV$ and $\mu =3 \GeV$, we obtain:
\begin{eqnarray} \label{eq: step_scale_results}
\sigma^{(\gamma^{\mu}, \gamma^{\mu})}_{(27,1)} (\mu, \mu_0) &=& 0.947\,(05)(01)\,, \\ 
[3mm]
\sigma^{(\gamma^{\mu}, \gamma^{\mu})}_{(8,8)} (\mu, \mu_0) &=&
\begin{pmatrix}
0.963\,(06)(14)\quad	&	0.376\,(16)(75) 	\\
0.040\,(19)(20)\quad	&	2.174\,(73)(91) 	\\
\end{pmatrix} \,,\\ 
[3mm]
\sigma^{(\qslashs, \qslashs)}_{(27,1)} (\mu, \mu_0) &=& 0.881\,(07)(07)\,, \\
[3mm]
\sigma^{(\qslashs, \qslashs)}_{(8,8)} (\mu, \mu_0) &=&
\begin{pmatrix}
0.970\,(08)(08)\quad	&	0.288\,(15)(55) 	\\
0.156\,(37)(46)\quad	&	1.855\,(81)(36) 	\\
\end{pmatrix} \,.
\end{eqnarray}
The first quoted errors are statistical, while the second are the systematic from the continuum extrapolation.
We remind the reader that these results are universal, they depend on the choice 
of scheme and on the number of flavours (here $n_f=3$) but not on the details
of the lattice implementation.
\pagebreak

\section{Results in $\msbar$}
\label{sec:msbar}
The matching to $\msbar$ is performed at the scale $\mu = 3 \GeV$ 
where perturbation theory is expected to converge rather well.
With $\alpha^{\overline{\rm{MS}}}_{s}(3 \GeV) = 0.24544$, 
the matching factors are~\cite{Lehner:2011fz}:
\begin{eqnarray}
C^{\msbar\leftarrow(\gamma^{\mu}, \gamma^{\mu})}_{(27,1)} (3 \GeV) &=& 1.00414 \\
[3mm]
C^{\msbar\leftarrow(\gamma^{\mu}, \gamma^{\mu})}_{(8,8)} (3 \GeV) &=& 
\begin{pmatrix}
\ph{-}1.00084	&	-0.00253 	\\
-0.03152 &	\ph{-}1.08781 	\\
\end{pmatrix} \\
[3mm]
C^{\msbar\leftarrow(\qslashs, \qslashs)}_{(27,1)}(3 \GeV) &=& 0.99112 \\
[3mm]
C^{\msbar\leftarrow(\qslashs, \qslashs)}_{(8,8)}(3 \GeV) &=& 
\begin{pmatrix}
\ph{-}1.00084	&	-0.00253	\\
-0.01199	&	\ph{-}1.02921	\\
\end{pmatrix} \,.
\end{eqnarray}
Finally the $Z$-factors in $\msbar$ 
are given by
\be
Z^{\msbar}(\mu) = 
C^{\msbar\leftarrow\schemeS} (\mu) \times \sigma^{\schemeS}(\mu,\mu_0) \times Z^{\schemeS}(\mu_0) 
\;,
\ee
we obtain:
\begin{eqnarray} 
\label{eq:Z_MSbar_results1}
 Z_{(27,1)}^{\msbar\leftarrow(\gamma^{\mu}, \gamma^{\mu})}(3\GeV)  &=&0.421\,(02)(00)\\ 
\label{eq:Z_MSbar_results2}
 Z_{(8,8)}^{\msbar\leftarrow(\gamma^{\mu}, \gamma^{\mu})}(3\GeV)  &=&
\begin{pmatrix}
\ph{-}0.479\,(03)(07)	&	-0.024\,(04)(17) 	\\
-0.045\,(11)(11)	&	\ph{-}0.543\,(18)(23) 	\\
\end{pmatrix} \\ [3mm]
\label{eq:Z_MSbar_results3}
 Z_{(27,1)}^{ \msbar\leftarrow(\qslashs,\qslashs)}(3\GeV)  &=& 0.427\,(03)(03) \\
\label{eq:Z_MSbar_results4}
 Z_{(8,8)}^{ \msbar\leftarrow(\qslashs,\qslashs)}(3\GeV)  &=& 
\begin{pmatrix}
\ph{-}0.473\,(05)(06)	&	-0.026\,(05)(17) 	\\
-0.070\,(23)(25)	&	\ph{-}0.564\,(27)(13) 	\\
\end{pmatrix} \,,
\end{eqnarray}
where the first quoted error combines statistical errors, while the second is due to
the continuum extrapolation systematic in~(\ref{eq:ssm}). 
In Eqs.~(\ref{eq:Z_MSbar_results1}-\ref{eq:Z_MSbar_results2}-\ref{eq:Z_MSbar_results3}-\ref{eq:Z_MSbar_results4}),
the superscript $\msbar\leftarrow\schemeS$ reminds us that the $Z$ factors were first 
evaluated in the scheme $\schemeS$. 
In principle the results should agree once converted in $\msbar$, but 
in practice they might be a difference due to lattice artefacts from 
the IDSDR lattice and due to the truncation of perturbation theory
in the conversion to $\msbar$. This difference can be use to estimate
the systematics errors or the renormalization procedure.


\section{Conclusion}
\label{sec:conclusion}
We have computed the matching factors required for the 
determination of the RBC-UKQCD collaborations' recently completed
calculation of $K \rightarrow \pi\pi$ decays in the $\Delta I = \frac{3}{2}$ channel.
The low energy matrix elements were computed on a single,
coarse ($a \sim 0.14 \text{ fm}$) IDSDR ensemble.
In order to make an end-run around the upper limit of the Rome-Southampton window,
we have calculated non-perturbative step-scaling functions on IW lattices with smaller
lattice spacings ($a \sim 0.114 \text{ fm and } 0.086 \text{ fm}$), and extrapolated them to the 
continuum. 
The use of twisted boundary conditions is a crucial element that makes this possible.
The end result is that we can apply the one-loop perturbative SMOM$\rightarrow \msbar$
matching factors at a scale 3 GeV where perturbation theory converges rather well.
The use of multiple intermediate schemes gives an additional useful handle on the
effect of truncation at one loop.
In the future we plan to apply the same strategy to the %
$\Delta I =1/2$ operators, which require the computation of eye-diagrams.
A complete computation of the relevant matrix element for unphysical kinematics
has been recently published in~\cite{Blum:2011pu} and reported at this conference
\cite{Liu:2011jp}. 
\vspace{1.cm}\\
{\bf Acknowledgements} \\

We warmly thank all our colleagues of the RBC-UKQCD collaborations, and in particular 
Norman Christ and Chris Sachrajda for suggestions and stimulating discussions, 
Rudy Arthur and Chirs Kelly for their help at different stages of the project.
N.G. is supported by STFC grant ST/G000522/1. A.L. is supported by STFC Grant ST/J000396/1.
We thank the STFC funded DiRAC facility (supported by STFC grant ST/H008845/1), 
the University of Southampton's Iridis cluster (supported by STFC grant ST/H008888/1)
and the EU grant 238353 (STRONGnet).

\bibliography{Z_ss}{}

\begin{thebibliography}{10}

\bibitem{Blum:2011ng}
T.~Blum, P.A. Boyle, N.H. Christ, N.~Garron, E.~Goode, et~al.
\newblock {The $K\to(\pi\pi)_{I=2}$ Decay Amplitude from Lattice QCD}.
\newblock 2011.

\bibitem{bob:lat11}
R.~D. Mawhinney.
\newblock \pos{Pos(Lattice 2011) 024}, 2011.

\bibitem{kelly:lat11}
C.~Kelly.
\newblock \pos{PoS(Lattice 2011) 285}, 2011.

\bibitem{Goode:2011kb}
Elaine~J. Goode and Matthew Lightman.
\newblock {$\Delta I = 3/2, K to \pi \pi$ Decays with a Nearly Physical Pion
  Mass}.
\newblock \pos{PoS(Lattice 2010) 313}, 2010.

\bibitem{Goode:2011pd}
Elaine Goode and Matthew Lightman.
\newblock {Delta I=3/2 K to pi-pi decays with nearly physical kinematics}.
\newblock \pos{PoS(Lattice 2011) 335}, 2011.

\bibitem{Martinelli:1994ty}
G.~Martinelli, C.~Pittori, Christopher~T. Sachrajda, M.~Testa, and A.~Vladikas.
\newblock {A General method for nonperturbative renormalization of lattice
  operators}.
\newblock {\em Nucl. Phys.}, B445:81--108, 1995.

\bibitem{Arthur:2010ht}
R.~Arthur and P.A. Boyle.
\newblock {Step Scaling with off-shell renormalisation}.
\newblock {\em Phys.Rev.}, D83:114511, 2011.

\bibitem{Arthur:2011cn}
R.~Arthur, P.A. Boyle, N.~Garron, C.~Kelly, and A.T. Lytle.
\newblock {Opening the Rome-Southampton window for operator mixing matrices}.
\newblock 2011.

\bibitem{Lehner:2011fz}
Christoph Lehner and Christian Sturm.
\newblock {Matching factors for Delta S=1 four-quark operators in RI/SMOM
  schemes}.
\newblock 2011.

\bibitem{Aoki:2010pe}
Y.~Aoki, R.~Arthur, T.~Blum, P.A. Boyle, D.~Brommel, et~al.
\newblock {Continuum Limit of $B_K$ from 2+1 Flavor Domain Wall QCD}.
\newblock 2010.

\bibitem{Gockeler:1998ye}
M.~Gockeler et~al.
\newblock {Nonperturbative renormalisation of composite operators in lattice
  QCD}.
\newblock {\em Nucl. Phys.}, B544:699--733, 1999.

\bibitem{Aoki:2007xm}
Y.~Aoki et~al.
\newblock {Non-perturbative renormalization of quark bilinear operators and
  $B_K$ using domain wall fermions}.
\newblock {\em Phys. Rev.}, D78:054510, 2008.

\bibitem{Babich:2006bh}
R.~Babich, N.~Garron, C.~Hoelbling, J.~Howard, L.~Lellouch, and C.~Rebbi.
\newblock {K0 - anti-K0 mixing beyond the standard model and CP-violating
  electroweak penguins in quenched QCD with exact chiral symmetry}.
\newblock {\em Phys.Rev.}, D74:073009, 2006.

\bibitem{Boyle:2011kn}
P.~Boyle and N.~Garron.
\newblock {Non-perturbative renormalization of kaon four-quark operators with
  nf=2+1 Domain Wall fermions}.
\newblock \pos{ PoS(Lattice 2010) 307}, 2010.

\bibitem{Dimopoulos:2010wq}
P.~Dimopoulos et~al.
\newblock {$K^0-\bar{K}^0$ Mixing Beyond the SM from Nf=2 tmQCD}.
\newblock \pos{ PoS(Lattice 2010) 302}, 2010.

\bibitem{Wennekers:2008sg}
J.~Wennekers.
\newblock {Neutral Kaon Mixing Beyond the Standard Model from 2+1 Flavour
  Domain Wall QCD}.
\newblock \pos{ PoS(Lattice 2008) 269}, 2008.

\bibitem{Aoki:2010dy}
Y.~Aoki et~al.
\newblock {Continuum Limit Physics from 2+1 Flavor Domain Wall QCD}.
\newblock 2010.

\bibitem{Allton:2008pn}
C.~Allton et~al.
\newblock {Physical Results from 2+1 Flavor Domain Wall QCD and SU(2) Chiral
  Perturbation Theory}.
\newblock {\em Phys.Rev.}, D78:114509, 2008.

\bibitem{Blum:2011pu}
T.~Blum, P.A. Boyle, N.H. Christ, N.~Garron, E.~Goode, et~al.
\newblock {$K$ to $\pi\pi$ Decay amplitudes from Lattice QCD}.
\newblock 2011.

\bibitem{Liu:2011jp}
Qi~Liu.
\newblock {Practical methods for a direct calculation of $\Delta I=1/2$ $K$ to
  $\pi\pi$ Decay}.
\newblock \pos{PoS(Lattice 2011) 288}, 2011.

\end{thebibliography}
\bibliographystyle{h-elsevier}

\end{document}